\documentclass[11pt]{article}
\usepackage[a4paper]{geometry}
\usepackage{amsmath}
\usepackage{amsfonts}
\usepackage{amsbsy}
\usepackage{amsthm}
\usepackage{accents}
\usepackage{tensor}
\usepackage{authblk}
\usepackage{braket}
\usepackage{graphicx}
\usepackage{caption}
\usepackage{subcaption}
\usepackage{cite}
\usepackage{color}
\usepackage{comment}
\usepackage{cancel}

\newcommand{\bs}[1]{\mathbf{#1}}

\newcommand{\kn}{|\vec{k}|}
\newcommand{\ofk}{(\vec{k})}
\newcommand{\volel}{\frac{d^3 k}{(2 \pi)^3 \kn}}
\newcommand{\rtt}{\widetilde{\mathbb{R}}^3}

\newcommand{\tPsi}{\widetilde{\Psi}}
\newcommand{\tPhi}{\widetilde{\Phi}}

\newcommand{\hrm}{\dag}
\newcommand{\be}{\begin{equation}}
\newcommand{\ee}{\end{equation}}

\theoremstyle{definition}

\newtheorem{rmrk}{Remark}

\numberwithin{equation}{section}

\begin{document}
\title{\bf Construction of a photon position operator with commuting components from natural axioms}  
\date{\today} 
\author[1]{Michał~Dobrski\thanks{michal.dobrski@p.lodz.pl}}
\author[1]{Maciej~Przanowski\thanks{Professor emeritus}\thanks{maciej.przanowski@p.lodz.pl}}
\author[1]{Jaromir~Tosiek\thanks{jaromir.tosiek@p.lodz.pl}}
\author[2]{Francisco~J.~Turrubiates\thanks{fturrubiatess@ipn.mx}}
\affil[1]{Institute of Physics, Łódź University of Technology,\linebreak Wólczańska 217/221, 93-005 Łódź, Poland}
\affil[2]{Departamento de Física, Escuela Superior de Física y Matemáticas, Instituto~Politécnico~Nacional, Unidad Adolfo López Mateos, Edificio 9, 07738~Ciudad~de~México, México} 
\maketitle 
\begin{abstract}
A general form of the photon position operator with commuting components fulfilling some natural axioms is obtained. This operator commutes with the photon helicity operator, is Hermitian with respect to the ~Białynicki -- Birula scalar product and defined up to a unitary transformation preserving the transversality condition. 
It is shown that, using the procedure analogous to the one introduced by T.~T.~Wu and C.~N.~Yang for the case of the Dirac magnetic monopole, the photon position operator can be defined by a flat connection in some trivial vector bundle over $\mathbb{R}^3 \setminus \{(0,0,0)\}$. This observation enables us to reformulate quantum mechanics of a~single photon on $(\mathbb{R}^{3} \setminus \{(0,0,0)\}) \times \mathbb{C}^2$.
\end{abstract}

\section{Introduction}

The question whether a photon position operator exists is a long standing and still fascinating problem in relativistic quantum mechanics. In 1948 M.~H.~L.~Pryce in his distinguished work \cite{pryce} on the mass -- centre in relativistic field theory defined an operator which he interpreted as the photon position operator. However, the components of that operator do not commute and, consequently, Pryce's operator does not fulfil the standard quantum mechanics requirements for the position operator.  T.~D.~Newton and E.~P.~Wigner \cite{newton} formulated a concept and derived several properties of localized states of elementary systems. They showed that for massless particles of spin equal or greater than one localized states satisfying the axioms assumed in \cite{newton} do not exist. The authors conclude that `this is an unsatisfactory, if not unexpected feature' of their work. According to this result there are no localized states for the photon.

A.~S.~Wightman \cite{wightman} reformulated the ideas of Newton and Wigner in terms of a set of projective operators constituting a {\it system of imprimitivity}. This set satisfies five axioms and it enabled Wightman to define commuting coordinate operators. Then Wightman showed that in the case of photon his method does not work and, therefore, the photon is not a localizable particle.

As shown by J.~M.~Jauch and C.~Piron \cite{jauch} and independently by W.~O.~Amrein \cite{amrein} one can weaken the Wightman axioms to get a {\it generalized system of imprimitivity} which enables one to introduce a weak localizability of the photon. From the point of view of quantum mechanical formalism the notion of the {\it photon localizability} is closely related (if not equivalent) to the notion of the photon position operator. This photon position operator, like the Pryce one, has noncommuting components.

The photon position operators with nonabelian components were analysed in detail by B.~S.~Skagerstam \cite{skagerstam, skagerstam1, skagerstam2}, and recently by P.~Kosi\'{n}ski and P.~Ma\'{s}lanka \cite{kosinski}.

On the other hand in 1999 a very promising photon position operator with commuting components was found by M.~Hawton \cite{hawton}. This idea was then developed in several publications \cite{hawton1, hawton2, hawton3,debierre, hawton4, hawton5,babaei,dobrski}. Significant progress in understanding Hawton's photon position operator was made in \cite{hawton1, hawton5}, where it was shown that one can construct other Hawton -- like operators by choosing a suitable orthonormal basis in a 3--D Euclidean momentum space. Very recently \cite{dobrski} we developed the idea presented in  articles  \cite{hawton1, hawton5} and  proposed a simple geometrical interpretation of the Hawton -- like photon position operators as objects determined by a flat connection (covariant derivative) on an appropriate dense subset of the momentum space ${\mathbb R}^3$ such that $2$--planes orthogonal to the momentum vector are propagated parallel, and moreover, the covariant differentiation is an antihermitian operator with respect to the  Białynicki -- Birula scalar product \cite{bialynicki, bialynicki1}. Note that similar axioms have been assumed in the work by H.~Babaei and A.~Mostafazedeh \cite{babaei}.

In \cite{dobrski} we also found eigenfunctions of the photon position operator in the coordinate representation and demonstrated that the behaviour of these functions is consistent with the original interpretation by I.~Białynicki -- Birula \cite{bialynicki, bialynicki1} and J.~E.~Sipe \cite{sipe}.

The present work can be considered as further development of theory of the photon position operator with commuting components. We follow the well known Dirac path used by him to build the momentum operator in the Schrödinger representation \cite{dirac}. 

In the case of photon we use the momentum representation. We assume that the canonical operator of the photon momentum is given and we look for the canonical photon position operator i.e.\ an operator which satisfies the canonical commutation relations, commutes with the photon helicity operator, whose action on a photon wave function transforms it into another photon wave function and which is Hermitian with respect to the Białynicki -- Birula scalar product. 

Thus in Sec.\ \ref{sec2} we find a general form of the photon position operator obeying the conditions mentioned above. The main result of that section is that the Hawton photon position operator as well as other   photon position operators with commuting components obtained in \cite{hawton1, hawton5, dobrski} are, up to a unitary transformation preserving the transversality condition \eqref{2.4}, the {\it general  photon position operators with commuting components}.

However, as it can be easily observed \cite{hawton,hawton1, hawton5, dobrski}, if one tries to express any of these operators globally on all the momentum space $\mathbb{R}^3 \setminus  \{(0,0,0)\}$, then some functions revealing string singularity or discontinuity appear. This feature of the photon position operator has been interpreted in the spirit of the work of T.~T.~Wu and C.~N.~Yang \cite{wu} on ``nonintegrable'' phase factors by M.~Hawton and W.~E.~Baylis in \cite{hawton1}. They have shown that the singular term of the photon position operator can be brought to a nonsingular form by an appropriate gauge transformation (of the form (\ref{2.26}) in the present paper). However the resulting expression reveals discontinuity, thus it is ``nonintegrable'' in the sense of \cite{wu}. This suggests that one can follow the path paved by T.~T.~Wu and C.~N.~Yang for the case of the Dirac magnetic monopole \cite{wu,wu2} and try to apply the vector bundle machinery to define the photon position operator. In our approach described in Sec.\ 3 we assume that the photon wave functions are sections of some bundle over $M={\mathbb R}^3 \setminus  \{(0,0,0)\}.$ This bundle turns out to be trivial and the photon position operator with commuting components is given by some flat connection in it multiplied by $\sqrt{-1}$. Using the trivialization isomorphism we show that the considered model of quantum mechanics of photon takes particularly simple form when it is transported to the explicitly trivial bundle $M \times \mathbb{C}^2$.
Concluding remarks end the paper.


\section{General form of the photon position operator with commuting components}
\label{sec2}

At the beginning of this section we recall some basic facts about quantum theory of a single photon.

The Hilbert space of photon states arises  from the tensor product of vector spaces $L^2_{(BB)}({\mathbb R}^3) \otimes {\mathbb C}^3$, where $L^2_{(BB)}({\mathbb R}^3)$ is the space of functions defined on ${\mathbb R}^3$, which are square integrable with respect to the relativistic invariant scalar product $(\psi,\phi)_{BB}=\int \volel \bar{\psi} \ofk \phi \ofk$.
Let~$\bs{1}$~denote the unit $3 \times 3$ matrix. 
All calculations are done exclusively in the canonical 
momentum representation. Thus the action of the 
photon momentum operator  
\be
\label{2.0}
  \hat{\vec p} \otimes \bs{1} = (\hat{p}_1 \otimes \bs{1},\hat{p}_2  \otimes \bs{1}, \hat{p}_3  \otimes \bs{1} )
  \ee
   is represented by multiplication by components of vector 
$\vec{p}=(p_1,p_2,p_3). $ Analogously the photon wave vector operator 
\be
\label{2.01}
\hat{\vec k} \otimes \bs{1}:= \frac{\hat{\vec p} \otimes \bs{1}}{\hbar}= (\hat{k}_1 \otimes \bs{1},\hat{k}_2 \otimes \bs{1}, \hat{k}_3 \otimes \bs{1} )
\ee
 action is realised by multiplication by numbers $k_1, k_2, k_3$ respectively. The spin-$1$ matrices are of the form
\[
\vec{\bs{S}}=(\bs{S}_1,\bs{S}_2,\bs{S}_3)
\]
\begin{align}
\label{2.1}
\bs{S}_1 &= 
\begin{pmatrix}
0 & 0 & 0\\
0 & 0 & -i\\
0 & i & 0 
\end{pmatrix},
&
\bs{S}_2 &= 
\begin{pmatrix}
0 & 0 & i\\
0 & 0 & 0\\
-i & 0 & 0 
\end{pmatrix},
&
\bs{S}_3 &= 
\begin{pmatrix}
0 & -i & 0\\
i & 0 & 0\\
0 & 0 & 0 
\end{pmatrix}.
\end{align}
The photon helicity operator is given by the matrix
\begin{equation}
\label{2.2}
\bs{\Sigma} = \frac{\vec{k} \cdot \vec{\bs{S}}}{\kn} =  i \kn^{-1}  
\begin{pmatrix}
0 & -k_3 & k_2\\
k_3 & 0 & -k_1\\
-k_2 & k_1 & 0 
\end{pmatrix}.
\end{equation}
We demand that  components of the canonical photon position operator $\hat{X}_j, \; j=1,2,3,$ satisfy the following axioms:
\begin{enumerate}
\item $[\hat{X}_j, \hat{X}_l] = 0,$ \label{ax1}
\item
$[\hat{X}_j, \hat{k}_l] = i \delta_{jl}, \;\;\; j,l=1,2,3,$ \label{ax2}
\item
$[\hat{X}_j, \bs{\Sigma}]= 0, $ \label{ax3}
\item
$ k_l \cdot  \left( \hat{X}_j \bs{\tPsi} \ofk \right)_l=0 $ for any \label{ax4}
\begin{equation}
\label{2.3}
\bs{\tPsi} \ofk = 
\begin{pmatrix}
\tPsi_1 \ofk \\
\tPsi_2 \ofk \\
\tPsi_3 \ofk 
\end{pmatrix}
\end{equation}
orthogonal to $\vec{k}$
\be
\label{2.4}
k_l \cdot {\tPsi}_l \ofk =0,
\ee
where summation over repeated indices from $1$ to $3$ is assumed.
\item \label{ax5}
Operators $ \hat{X}_j, \; j=1,2,3,$ are Hermitian with respect to the  Białynicki -- Birula scalar product 
\begin{equation}
\label{2.5}
\langle \bs{\tPhi} | \bs{\tPsi} \rangle_{\mathrm{BB}} := \int \volel \bs{\tPhi}^{\hrm} \ofk \bs{\tPsi} \ofk.
\end{equation}
\end{enumerate}

\begin{rmrk}
In this paper we assume that a photon wave function in the momentum representation is of the form \eqref{2.3} 
\cite{sipe, przanowski} and is orthogonal to $\vec{k}$, i.e.\ it obeys the transversality condition \eqref{2.4}. The motivation for such a choice of a photon wave function and relevant details can be found in \cite{przanowski}. Here, for readers concerned in the dynamics of the considered model, we briefly recall the following essential facts. 
The projections on the eigenspaces of the helicity operator $\bs{\Sigma}$ with helicities $+1$ and $-1$ 
\begin{align*}
\hat{\Pi}_+ \bs{\tPsi} \ofk &= \bs{\tPsi}_+ \ofk,\\
\hat{\Pi}_- \bs{\tPsi} \ofk &= \bs{\tPsi}_- \ofk
\end{align*}
give the unique decomposition of an arbitrary wave function \eqref{2.3} satisfying \eqref{2.4} into a sum
\begin{equation*}
\bs{\tPsi} \ofk =\bs{\tPsi}_+ \ofk + \bs{\tPsi}_- \ofk
\end{equation*}
which can be viewed as a superposition of two possible photon polarizations. The dynamics of both polarizations is handled by the single Schrödinger equation
\begin{equation*}
i \hbar \partial_t \bs{\tPsi}=\hat{H} \bs{\tPsi},
\end{equation*}
with the Hamiltonian given by
\begin{equation*}
\hat{H} = c \big(\vec{\bs{S}} \cdot \hat{\vec{p}}\big) \big(\hat{\Pi}_+ - \hat{\Pi}_- \big)
\end{equation*}
Reference \cite{przanowski} justifies this dynamics by an analogy to a certain formulation of Maxwell equations. In the present paper we work solely in the momentum representation, thus the definition and properties of helicity operator can be used to bring above Hamiltonian to 
\begin{equation*}
\hat{H} = c \hbar |\vec{k}| \bs{\Sigma} \left(\hat{\Pi}_+ - \hat{\Pi}_- \right)  
= c \hbar |\vec{k}| \left(\hat{\Pi}_+ + \hat{\Pi}_- \right) = c \hbar |\vec{k}|
\end{equation*}
yielding the following simple form of Schrödinger equation
\begin{equation*}
i \partial_t \bs{\tPsi} \ofk= c |\vec{k}| \bs{\tPsi} \ofk.
\end{equation*}
\end{rmrk}

Axioms {\ref{ax1}} and {\ref{ax2}}     
simply state that the photon position operator $\hat{\vec{X}} = (\hat{X}_1,\hat{X}_2,\hat{X}_3)$ and the momentum operator \eqref{2.0} satisfy the usual commutation relations.


The requirement {\ref{ax4}} says that operators $\hat{X}_j, \; j=1,2,3$ acting on any photon state turn it into another photon state. Axiom {\ref{ax5}} states that photon position operator 
$\hat{\vec{X}}$ is an observable with respect to the Białynicki -- Birula scalar product \eqref{2.5}. 

Then, from the postulate {\ref{ax3}} under {\ref{ax2}} it follows that position operator  $\hat{\vec{X}}$ and spin-$1$ matrices $\vec{\bs S}$ do not commute. Thus components of the photon total angular momentum operator chosen as
\be
\label{2.6}
\hat{J}_j= \varepsilon_{jlm} \hat{X}_l \hat{p}_m + \hbar \bs{S}_j, \;\;\; j=1,2,3
\ee
and the components of photon position operator $\hat{X}_j$ do not satisfy the standard commutation relations
\[
[\hat{X}_j, \hat{J}_l]= - i \hbar \varepsilon_{jlm} \hat{X}_m
\]
but the following ones
\be
\label{2.7}
[\hat{X}_j, \hat{J}_l]= - i \hbar \varepsilon_{jlm} \hat{X}_m + \hbar[\hat{X}_j,\bs{S}_l ].
\ee
Moreover, we should note that in fact the operator (\ref{2.6}) is not a ``photon operator'' since it does not transform the photon function again into a photon function. This is a result of the fact that it is not the spin $\vec{\bs{S}}$ but rather helicity $\bs{\Sigma}$ that has a meaning of the actual physical degree of freedom of a photon. One quickly finds that the photon orbital angular momentum operator $\hat{\vec{L}}= \hat{\vec{X}} \times \hat{\vec{p}}$ and the position operator $\hat{\vec{X}}$ fulfil the standard commutation relations
\begin{equation}
[\hat{X}_j, \hat{L}_l]= - i \hbar \varepsilon_{jlm} \hat{X}_m.
\end{equation}
Note that axiom {\ref{ax3}} together with {\ref{ax2}} and \eqref{2.2} yield the  relations
\be
\label{2.8}
k_l [\hat{X}_j,\bs{S}_l ]= i \left( \frac{k_j}{|\vec{k}|} \bs{\Sigma} - \bs{S}_j \right).
\ee
Pryce \cite{pryce}, Skagerstam \cite{skagerstam1, skagerstam2}, Kosi\'{n}ski and Ma\'{s}lanka \cite{kosinski} considered the photon  position operator satisfying the standard commutation relations with the photon angular momentum but this assumption leads to a photon position operator with noncommuting components.

Our idea is to use postulates {\ref{ax1}} -- {\ref{ax5}} to find a general form of the photon position operator with commuting components. 
Let operators $\hat{\vec{X}} = (\hat{X}_1,\hat{X}_2,\hat{X}_3)$ and $\hat{\vec{X}}' = (\hat{X}_1',\hat{X}_2',\hat{X}_3')$
obey axioms  {\ref{ax1}} -- {\ref{ax5}}. One can write
\be
\label{2.9}
\hat{X}_j' = \hat{X}_j + \hat{Y}_j, \;\;\; j=1,2,3.
\ee
Since both operators  $\hat{\vec{X}}$ and  $\hat{\vec{X}}'$ fulfil conditions {\ref{ax2}} and {\ref{ax3}}, we obtain the following commutation rules
\be
\label{2.10}
[\hat{Y}_j, \hat{k}_l]= \hat{0} \;\;\; , \;\;\; [\hat{Y}_j, \bs{\Sigma}]=0, \;\;\; j,l=1,2,3.
\ee
It is well known that components of the photon wave vector operator, $\hat{k}_1, \hat{k}_2, \hat{k}_3 $, constitute a complete set of commuting observables acting in the Hilbert space $L^2_{(BB)}({\mathbb R}^3)$. Therefore operator $\hat{Y}_j$ is represented by a matrix
\be
\label{2.11}
\bs{Y}_j=\begin{pmatrix}
y_{j,11} & y_{j,12} & y_{j,13}\\
y_{j,21} & y_{j,22} & y_{j,23}\\
y_{j,31} & y_{j,32} & y_{j,33}
\end{pmatrix}, \;\;\; j=1,2,3
\ee
where $y_{j,lm}=y_{j,lm}(\vec{k}) $ are functions of $\vec{k}=(k_1, k_2, k_3).$ 

Eigenvalues of the photon helicity operator $\bs{\Sigma}$ are $\{-1,1,0\}$, so its spectrum is not degenerated. Then by some unitary (with respect to standard and to the Białynicki -- Birula scalar product) transformation the matrix $\bs{\Sigma}$ can be brought to a diagonal form
\be
\label{2.12}
\bs{\Sigma}=
\begin{pmatrix}1 & 0 & 0 \\
0 & -1 & 0 \\
0& 0 & 0
\end{pmatrix}.
\ee
Since operators $\hat{Y}_j$  and $\bs{\Sigma}$ commute for every  index $j=1,2,3$ then in the respective representation the
matrix  $\bs{Y}_j$ takes the form
\be
\label{2.14}
\bs{Y}_j=\begin{pmatrix}
y_{j,11} &0 & 0\\
0 & y_{j,22} & 0\\
0 & 0 & y_{j,33}
\end{pmatrix}, \;\;\; j=1,2,3
\ee
where $y_{j,11}=y_{j,11}(\vec{k}), y_{j,22}=y_{j,22}(\vec{k}), y_{j,33}=y_{j,33}(\vec{k})$ are arbitrary but, by virtue of axiom {\ref{ax5}}, real functions of wave vector $\vec{k}.$

Using \eqref{2.12} one quickly finds that  matrix $\bs{Y}_j$ in the form \eqref{2.14} can be written as
\be
\label{2.15}
\bs{Y}_j = y_{j,33}\bs{1} + \frac{1}{2} \left( y_{j,11}- y_{j,22}\right) \bs{\Sigma} +
 \frac{1}{2} \left( y_{j,11}+ y_{j,22}-2 y_{j,33}\right) \bs{\Sigma}^2. 
\ee
Therefore in any basis of the Hilbert space ${\mathbb C}^3$
\be
\label{2.16}
\bs{Y}_j = \alpha_j(\vec{k})\bs{1} + \beta_j(\vec{k}) \bs{\Sigma} +
\gamma_j(\vec{k}) \bs{\Sigma}^2, \;\;\; j=1,2,3, 
\ee
where $\alpha_j(\vec{k}), \beta_j(\vec{k}) , \gamma_j(\vec{k})$ are real functions of wave vector $\vec{k}.$

Since both photon position operators  $\hat{\vec{X}}$ and  $\hat{\vec{X}}'$ fulfil requirements {\ref{ax1}} -- {\ref{ax5}}, from the first postulate {\ref{ax1}} one deduces that
\be
\label{2.17}
[\hat{X}_j', \hat{X}_l']= [\hat{X}_j+\bs{Y}_j, \hat{X}_l+\bs{Y}_l]= 
 [\hat{X}_j, \bs{Y}_l]- [\hat{X}_l, \bs{Y}_j]=0.
\ee
Inserting \eqref{2.16} into \eqref{2.17} and using axioms {\ref{ax2}} and {\ref{ax3}} we obtain
\be
\label{2.18}
\left(\frac{\partial \alpha_l}{\partial k_j} - \frac{\partial \alpha_j}{\partial k_l} \right) \bs{1}+
\left(\frac{\partial \beta_l}{\partial k_j} - \frac{\partial \beta_j}{\partial k_l} \right) \bs{\Sigma}+
\left(\frac{\partial \gamma_l}{\partial k_j} - \frac{\partial \gamma_j}{\partial k_l} \right) \bs{\Sigma}^2=0, \;\;\; j,l=1,2,3.
\ee
Hence
\be
\label{2.19}
\frac{\partial \alpha_l}{\partial k_j} - \frac{\partial \alpha_j}{\partial k_l}=0 \;\; ,\;\;
\frac{\partial \beta_l}{\partial k_j} - \frac{\partial \beta_j}{\partial k_l}=0 \;\;{\rm and}\;\;
\frac{\partial \gamma_l}{\partial k_j} - \frac{\partial \gamma_j}{\partial k_l}=0, \;\;\;  j,l=1,2,3.
\ee
Consequently for a simply connected domain, from \eqref{2.19}  we conclude that there exist real functions $A=A(\vec{k}), B=B(\vec{k})$ and $C=C(\vec{k})$ such that
\be
\label{2.20}
\alpha_j = \frac{\partial A}{\partial k_j}\;\; , \;\; \beta_j = \frac{\partial B}{\partial k_j} \;\; , \;\; 
\gamma_j = \frac{\partial C}{\partial k_j}.
\ee
Substituting \eqref{2.20} into \eqref{2.16} one gets a general form of $\bs{Y}_j$ as 
\be
\label{2.21}
\bs{Y}_j = \frac{\partial A}{\partial k_j}\bs{1} +  \frac{\partial B}{\partial k_j} \bs{\Sigma} +
\frac{\partial C}{\partial k_j} \bs{\Sigma}^2, \;\;\; j=1,2,3,
\ee 
where $A=A(\vec{k}), B=B(\vec{k})$ and $C=C(\vec{k})$ are any real functions of $\vec{k}.$ 

Gathering together our considerations we arrive at the conclusion that given any photon position operator with commuting components
 $\hat{\vec{X}} = (\hat{X}_1,\hat{X}_2,\hat{X}_3)$ a general form of the photon position operator
  $\hat{\vec{X}}' = (\hat{X}_1',\hat{X}_2',\hat{X}_3')$ reads
  \be
  \label{2.22}
  \hat{X}_j'= \hat{X}_j +\frac{\partial A}{\partial k_j}\bs{1} +  \frac{\partial B}{\partial k_j} \bs{\Sigma} +
\frac{\partial C}{\partial k_j} \bs{\Sigma}^2, \;\;\; j=1,2,3,
  \ee
 where $A=A(\vec{k}), B=B(\vec{k})$ and $C=C(\vec{k})$ are arbitrary real functions of $\vec{k}.$
 
 At this point we apply a procedure analogous to that one used by P.~A.~M.~Dirac in his epochal book \cite{dirac} to obtain a standard form of the momentum operator in the Schrödinger representation. Namely we perform the following unitary transformation in the sense of both the Białynicki -- Birula and the standard scalar product
 \[
 \hat{X_j}' \mapsto \hat{X_j}'' = \exp \left\{ -i \left( A(\vec{k}) + C(\vec{k}) \right) \right\} \hat{X_j}' \exp \left\{ i \left( A(\vec{k}) + C(\vec{k}) \right) \right\}
 \]
\be
\label{2.23}
 =\hat{X_j}+ \frac{\partial B(\vec{k})}{\partial k_j} \bs{\Sigma} +
\frac{\partial C(\vec{k})}{\partial k_j} \left( \bs{\Sigma}^2 - {\bs{1}} \right)
\ee
with the use of axiom {\ref{ax2}}. However, the operator \eqref{2.23} considered on the space of photon wave functions can be written in more compact form. Observe that by \eqref{2.2} 
\[
\left( \bs{\Sigma}^2 \right)_{jl} = \delta_{jl} - \frac{k_j k_l}{|\vec{k}|^2}
\]
so
\be
\label{2.24}
\left( \bs{\Sigma}^2 \right)^2 = \bs{\Sigma}^2 \;\; \;, \;\;\; \left( \bs{\Sigma}^2 \right)^{\dagger}= \bs{\Sigma}^2.
\ee
Therefore $\bs{\Sigma}^2$ is a projection operator in the space $L^2_{(BB)}({\mathbb R}^3) \otimes {\mathbb C}^3$ of functions
\be
\label{2.25}
\widetilde{\bs{\Phi}}(\vec{k}) = \begin{pmatrix}
\widetilde{\Phi}_1(\vec{k}) \\ \widetilde{\Phi}_2(\vec{k}) \\ \widetilde{\Phi}_3(\vec{k}) 
\end{pmatrix}
\ee
and it projects an arbitrary function $\widetilde{\bs{\Phi}}(\vec{k}) \in L^2_{(BB)}({\mathbb R}^3) \otimes {\mathbb C}^3$ onto
$\widetilde{\bs{\Psi}}(\vec{k}):=\bs{\Sigma}^2  \widetilde{\bs{\Phi}}(\vec{k})  $ fulfilling transversality condition \eqref{2.4}. Consequently, operator $\bs{\Sigma}^2$ restricted to the space of photon wave functions is equivalent to the identity operator $\bs{1}.$ It means that when operator $\hat{X_j}''$ acts on any photon wave function, the result is equal to the action of operator $\hat{X_j}+ \frac{\partial B(\vec{k})}{\partial k_j} \bs{\Sigma}$ on this function.

Thus we reach the following result. If  $\hat{\vec{X}}= (\hat{X}_1, \hat{X}_2, \hat{X}_3)$ is a photon position operator fulfilling requirements {\ref{ax1}} -- {\ref{ax5}} then every position operator $\hat{\vec{X}}'= (\hat{X}_1', \hat{X}_2', \hat{X}_3')$ obeying the same set of axioms {\ref{ax1}} -- {\ref{ax5}} is of the form \eqref{2.22}. Moreover, there exists  a unitary in the sense of both the Białynicki -- Birula and the standard scalar products transformation mapping the space of photon states onto the same space such that this transformation changes $\hat{\vec{X}}'$ into $\hat{\vec{X}}''$ according to rule \eqref{2.23}. On the space of photon wave functions satisfying the transversality condition \eqref{2.4} we can write
\be
\label{2.26}
\hat{\vec{X}}'' = \hat{\vec{X}}+ \left( \nabla_{\vec{k}} B(\vec{k})\right) \cdot \bs{\Sigma},
\ee
where $\nabla_{\vec{k}}= \left(\frac{ \partial}{\partial k_1}, \frac{ \partial}{\partial k_2}, \frac{ \partial}{\partial k_3}\right)$ and $ B(\vec{k})$ is a real function of $\vec{k}.$ (The term $ \nabla_{\vec{k}} B(\vec{k})$ is analogous to $\pmb{\nabla}\chi(\theta,\phi)$ considered in e.g.\ \cite{hawton5}. However, $B \ofk$ may depend not only on angular coordinates, but on the whole $\vec{k}$. It also appears as a consequence of our set of axioms, before any specific realization of $\hat{\vec{X}}$ is chosen.)

To proceed further we employ results of our recent work \cite{dobrski}. In that paper  we showed that the photon position operator with commuting components introduced by Hawton \cite{hawton} is determined by a flat connection in an appropriate open dense submanifold of momentum space ${\mathbb R}^3$ such that tangent planes orthogonal to the momentum are propagated parallel with respect to this connection and that a corresponding  covariant derivative is an anti -- Hermitian operator with respect to the Białynicki -- Birula scalar product \eqref{2.5}. 
The submanifold of $\mathbb{R}^3$ we consider in \cite{dobrski} is defined  as $\rtt := \mathbb{R}^3 \setminus \{ (0,0,k_3) \in \mathbb{R}^3: k_3 \geq 0 \}$. Here we use the same submanifold but, as it is explained in \cite{dobrski}, other dense submanifolds can be used equivalently. 
Then we demonstrated that a general  photon position operator $\hat{\vec{X}}$  constructed in this way acts on a vector function $ \widetilde{\bs{\Phi}}=\widetilde{\bs{\Phi}}(\vec{k})$ given by \eqref{2.25} so that on $\rtt$ 
\be
\label{2.27}
\left( \hat{X}_j \widetilde{\bs{\Phi}} \right)_l = i e_{\mu l} \frac{\partial e^{\mu}_m \widetilde{\Phi}_m}{\partial k_j}\;\;\;, \;\;\; j,l,m=1,2,3 \;\;\;, \;\;\;\mu=1,2,3
\ee
where
\be
\label{2.28}
\vec{e}_{\mu} = \vec{e}_{\mu}(\vec{k}\,) \;\;\; , \;\;\; \vec{e}_{\mu}= e_{\mu l} \frac{\partial}{\partial k_l} \;\;\;, \;\;\;\mu=1,2,3
\ee
are orthogonal basis vector fields on $\rtt$ such that $\vec{e}_{1}, \vec{e}_{2} \, \bot \,\vec{k},$
\be
\label{2.29}
\vec{e}^{\,\mu} = \vec{e}^{\,\mu}(\vec{k}) \;\;\; , \;\;\; \vec{e}^{\,\mu}= {e^{\mu}}_{ l} dk_l \;\;\;, \;\;\;\mu=1,2,3
\ee
are the basis $1$--forms dual to $\big(\vec{e}_{1}(\vec{k}), \vec{e}_{2}(\vec{k}), \vec{e}_{3}(\vec{k}) \big)$ i.e.
\be
\label{2.30}
\vec{e}^{\,\mu}\big(\vec{e}_{\nu} \big) = {\delta^{\mu}}_{\nu} \Longleftrightarrow {e^{\mu}}_{ l} e_{\nu l}= {\delta^{\mu}}_{\nu} \;\;\;, \;\;\;\mu=1,2,3.
\ee
Without any loss of generality one can assume that the basis $\big(\vec{e}_{1}, \vec{e}_{2}, \vec{e}_{3} \big)$ is right oriented and $\vec{e}_{3}(\vec{k})= \omega (\vec{k}) \vec{k}, \; \omega (\vec{k})>0.$ 

In fact, there exists a system of conditions determining the function $\omega(\vec{k}).$ Indeed, as we proved in \cite{dobrski}, the triad of vector fields $\left(\vec{E}_{1}(\vec{k}), \vec{E}_{2}(\vec{k}), \vec{E}_{3}(\vec{k}) \right)$  on $\rtt$ defined by
\be
\label{2.31}
\vec{E}_{\mu}= \vec{E}_{\mu}(\vec{k})=E_{\mu j} \frac{\partial}{\partial k_j}:= |\vec{k}|^{-1/2} \vec{e}_{\mu}(\vec{k}) \;\;\;, \;\;\;\mu=1,2,3
\ee
constitutes a right oriented orthonormal basis
\be
\label{2.32}
\vec{E}_{\mu}(\vec{k}) \cdot \vec{E}_{\nu}(\vec{k}) = \delta_{\mu \nu}\;\;\;, \;\;\;
\vec{E}_{1}(\vec{k}), \vec{E}_{2}(\vec{k}) \, \bot \,\vec{k} \;\;\;, \;\;\;
\vec{E}_{1}(\vec{k}) \times \vec{E}_{2}(\vec{k}) =  \vec{E}_{3}(\vec{k}) = \frac{\vec{k}}{|\vec{k}|}.
\ee
Basis of $1$--forms $\big(\vec{E}^{1}, \vec{E}^{2}, \vec{E}^{3}\big)$ dual to $\big(\vec{E}_{1}, \vec{E}_{2}, \vec{E}_{3} \big)$ reads
\be
\label{2.33}
\vec{E}^{\mu}=\vec{E}^{\mu}(\vec{k})={E^{\mu}}_j dk_j:=  |\vec{k}|^{1/2} \vec{e}^{\,\mu}(\vec{k}) \;\;\;, \;\;\;\mu=1,2,3.
\ee
It is evident that ${E^{\mu}}_j=E_{\mu j}.$ 

Inserting \eqref{2.31} and \eqref{2.33} into \eqref{2.27} and introducing a $3 \times 3$ matrix 
\be
\label{2.34}
(\bs{E})_{j  \mu}:= E_{\mu j} \;\;\;, \;\;\; \mu, j=1,2,3
\ee
(compare with Eq. (2.33) in \cite{dobrski}) one can rewrite \eqref{2.27} as
\be
\label{2.35}
\hat{X}_j \widetilde{\bs{\Phi}}  = i  |\vec{k}|^{1/2} \bs{E} \frac{\partial \left(
|\vec{k}|^{-1/2} 
\bs{E}^{-1} \widetilde{\bs{\Phi}}\right)}{\partial k_j}\;\;\;, \;\;\; j=1,2,3,
\ee
on $\rtt$, where we use the fact that since $\big(\vec{E}_{1}, \vec{E}_{2}, \vec{E}_{3}\big)$ is an orthonormal basis, the matrix $\bs{E}$ is orthogonal i.e. $\bs{E}^{-1}= \bs{E}^{\rm T}.$ The photon position operator given by \eqref{2.27} or, equivalently, by \eqref{2.35} satisfies all axioms {\ref{ax1}} -- {\ref{ax5} } for any orthonormal basis 
$\big(\vec{E}_{1}, \vec{E}_{2}, \vec{E}_{3}\big)$ fulfilling conditions \eqref{2.32}. An arbitrary basis satisfying these conditions reads
\be
\left\{
\begin{array}{l}
  \vec{E}''_1 = a \vec{E}_1 - b \vec{E}_2  \\
\vec{E}''_2 = b \vec{E}_1 +a \vec{E}_2 
 \\
 \vec{E}''_3 =  \vec{E}_3 \label{2.36}
\end{array}
\right.
\ee
where
\[
a=a(\vec{k})\; ,\;b=b(\vec{k})\; ,\; a^2 +b^2=1 \; , \; \vec{k} \in \rtt
\]
and 
$\big(\vec{E}_{1}, \vec{E}_{2}, \vec{E}_{3}\big)$ is a given basis satisfying conditions \eqref{2.32}.

Transformation \eqref{2.36} can be written in a matrix form as
\be
\label{2.37}
\bs{E}''=\bs{E} \cdot \bs{U},
\ee
\[
\bs{U}:=\begin{pmatrix}
a & b & 0 \\
-b & a & 0 \\
0& 0& 1
\end{pmatrix}, \;\; a=a(\vec{k})\; ,\;b=b(\vec{k})\; ,\; a^2 +b^2=1.
\]
Note that $\bs{U}$ is a real unitary matrix, $\bs{U}^{-1}=\bs{U}^{\rm T}. $
With the use of $\bs{E}''$ we define the photon position operator $\hat{\vec{X}}''$ according to \eqref{2.35}. Then employing \eqref{2.37} one obtains
\begin{align}
\label{2.38}
\hat{X}_j'' \widetilde{\bs {\Phi}}& = i |\vec{k}|^{1/2}\bs{E}'' \frac{\partial \left( |\vec{k}|^{-1/2} {\bs{E}''}^{-1} \widetilde{\bs \Phi}\right)}{\partial k_j} \nonumber \\
&= \hat{X}_j \widetilde{\bs \Phi} +i \bs{E} \left(\bs{U} \frac{\partial \bs{U}^{-1}}{\partial k_j} \right) \bs{E}^{-1} \widetilde{\bs{ \Phi}} \nonumber \\
& = \hat{X}_j \widetilde{\bs {\Phi}} + \left( a \frac{\partial b}{\partial k_j }-
b\frac{\partial a}{\partial k_j } 
\right) \bs{\Sigma} \widetilde{\bs {\Phi}}
\end{align}
on $\rtt$, where $\hat{X}_j$ is given by \eqref{2.35}. The exterior differential of equation $ a^2 +b^2=1$ leads to the observation that
\be
\label{2.39}
d( a^2 +b^2) = 2a da + 2b db=0 \Longrightarrow da \wedge db=0.
\ee
Define the following $1$--form 
\[
\sigma:= \left( a \frac{\partial b}{\partial k_j }-
b\frac{\partial a}{\partial k_j } 
\right) dk_j.
\]
Then by \eqref{2.39}
\be
\label{2.40}
d \sigma = 2 da \wedge db =0.
\ee
Since $\rtt$ is simply connected we infer from \eqref{2.40} that there exists a real function $B=B(\vec{k})$ such that
\be
\label{2.41}
\sigma= dB \Longrightarrow  a \frac{\partial b}{\partial k_j }-
b\frac{\partial a}{\partial k_j } = \frac{\partial B}{\partial k_j }.
\ee
Conversely, for any real function $B=B(\vec{k})$ we put
\be
\label{2.42}
a = a(\vec{k}):= \cos B(\vec{k}) \;\;\; , \;\;\;
b = b(\vec{k}):= \sin B(\vec{k}).
\ee
As one can see
\[
a^2 +b^2=1 \;\;\; {\rm and} \;\;\;  a \frac{\partial b}{\partial k_j }-
b\frac{\partial a}{\partial k_j } = \frac{\partial B}{\partial k_j }\;\;\; , \;\;\; j=1,2,3.
\]
Therefore, the general photon position operator  $\hat{X}_j''$ defined by \eqref{2.38} has the form
\be
\label{2.43}
\hat{X}_j''= i |\vec{k}|^{1/2}\bs{E}\, \frac{\partial  }{\partial k_j}  |\vec{k}|^{-1/2}\bs{E}^{-1}  + \frac{\partial B(\vec{k})}{\partial k_j } \bs{\Sigma}
\ee
where $\bs{E}$ is determined by some given basis according to \eqref{2.34} and $B= B(\vec{k})$ is an arbitrary real function of $\vec{k}.$  Thus we have reconstructed the formula \eqref{2.26} for $\hat{\vec{X}}$ given as
\be
\label{2.44}
\hat{\vec{X}}= i |\vec{k}|^{1/2}\bs{E} \; \nabla_{\vec{k}} \, |\vec{k}|^{-1/2}\bs{E}^{-1}.
\ee
Concluding, we arrive at an important result: {\it the general photon position operator \eqref{2.22} restricted to the linear space of photon wave functions satisfying the transversality condition \eqref{2.4} can be written as 
\begin{align}
\label{2.45}
\hat{\vec{X}}' & = \exp \left\{i F(\vec{k}) \right\}
i |\vec{k}|^{1/2}\bs{E}'' \; \nabla_{\vec{k}} \, |\vec{k}|^{-1/2}{\bs{E}''}^{-1}  \exp \left\{-i F(\vec{k}) \right\} \nonumber \\
& = \exp \left\{i F(\vec{k}) \right\}
i |\vec{k}|^{1/2}\bs{E} \; \nabla_{\vec{k}} \, |\vec{k}|^{-1/2}{\bs{E}}^{-1}  \exp \left\{-i F(\vec{k}) \right\} + \left( \nabla_{\vec{k}} B(\vec{k})\right)
\bs{\Sigma}
\end{align}
where $F=F(\vec{k})$ and $B=B(\vec{k})$ are arbitrary real functions of $\vec{k}.$}

A general form of the photon position operator satisfying conditions {\ref{ax1}} -- {\ref{ax5}} and acting on the linear space of $3$--vector functions \eqref{2.25} not necessarily satisfying transversality condition \eqref{2.4} is given by \eqref{2.22} and \eqref{2.23}
\begin{align}
\label{2.46}
\hat{\vec{X}}' & = \exp \left\{i F(\vec{k}) \right\}
i |\vec{k}|^{1/2}\bs{E}'' \; \nabla_{\vec{k}} \, |\vec{k}|^{-1/2}{\bs{E}''}^{-1}  \exp \left\{-i F(\vec{k}) \right\}
+\left( \nabla_{\vec{k}} C(\vec{k})\right)\left( \bs{\Sigma}^2- \bs{1} \right)
 \nonumber \\
& = \exp \left\{i F(\vec{k}) \right\}
i |\vec{k}|^{1/2}\bs{E} \; \nabla_{\vec{k}} \, |\vec{k}|^{-1/2}{\bs{E}}^{-1}  \exp \left\{-i F(\vec{k}) \right\} + \left( \nabla_{\vec{k}} B(\vec{k})\right) \bs{\Sigma} \nonumber \\
& \;\;\;\;+\left( \nabla_{\vec{k}} C(\vec{k})\right)\left( \bs{\Sigma}^2- \bs{1} \right).
\end{align}
Unitary transformation
\be
\label{2.231}
\exp \left\{i F(\vec{k}) \right\} [ \ldots ] \exp \left\{-i F(\vec{k}) \right\}
\ee
used in \eqref{2.45} and \eqref{2.46}
is obviously a counterpart of the unitary transformation considered by Dirac to turn the momentum operator in the Schrödinger representation to standard form $\hat{p}_r= - i \hbar \frac{\partial}{\partial q^r}$ (see \cite{dirac}, chapter 22).  Transformation \eqref{2.231} does not affect the momentum operator of photon, as well as the Hamilton operator or the helicity operator $\bs{\Sigma}.$ Moreover, it preserves transversality \eqref{2.4} of the photon wave function. Hence, without any loss of generality we put
\be
\label{2.47}
F(\vec{k})=0.
\ee
However, employing results presented in \cite{dobrski} we can proceed further. It is known that every simultaneous unitary transformation of operators and vectors preserves eigenvalues of operators, the time evolution of a system and the scalar product of vectors. Thus, if operators are unitarily equivalent, they can be also considered physically equivalent, provided that the unitary transformation is accordingly applied to all components of a theory. In \cite{dobrski} we showed that operators \eqref{2.38} and \eqref{2.44} are related by a unitary transformation (see Eq. (2.60) of \cite{dobrski}). We repeat that proof in a slightly different form  
\begin{align}
\label{2.48}
i |\vec{k}|^{1/2}\bs{E}'' \; \nabla_{\vec{k}} \, |\vec{k}|^{-1/2}{\bs{E}''}^{-1}=& 
i |\vec{k}|^{1/2}\bs{E}  \bs{U} \; \nabla_{\vec{k}} \, |\vec{k}|^{-1/2}\bs{U}^{-1}  {\bs{E}}^{-1} \nonumber \\
=& i |\vec{k}|^{1/2}\bs{E}  \bs{U} {\bs{E}}^{-1}  \bs{E}\; \nabla_{\vec{k}} \, |\vec{k}|^{-1/2}
{\bs{E}}^{-1}  \bs{E}
\bs{U}^{-1}  {\bs{E}}^{-1} \nonumber \\
=& \bs{V }  i |\vec{k}|^{1/2}\bs{E} \; \nabla_{\vec{k}} \, |\vec{k}|^{-1/2}{\bs{E}}^{-1}  \bs{V}^{-1},
\end{align}
where $\bs{V}:= \bs{E}  \bs{U}  \bs{E}^{-1}= \bs{E}''  \bs{E}^{-1}$ is a real unitary matrix $\bs{V}^{-1}= \bs{V}^{\rm T}.$
Moreover, one can check easily that unitary transformation
\[
\bs{V} [\ldots] \bs{V}^{-1}
\]
does not change the form of canonical momentum operator, the form of Hamilton operator and the form of operator $\bs{\Sigma}$ given by \eqref{2.2}. It also maintains the transversality property of the photon wave function
\be
\label{2.49}
(k_1\; k_2\;k_3) \cdot \widetilde{\bs{\Psi}}(\vec{k})=0 \Longleftrightarrow (k_1\; k_2\;k_3) \cdot \bs{V}  \widetilde{\bs{\Psi}}(\vec{k})=0.
\ee 
\emph{Therefore without any loss of generality one concludes that the general photon position operator satisfying axioms {\ref{ax1}} -- {\ref{ax5}}  and restricted to the space of photon wave functions fulfilling the transversality condition \eqref{2.4} can be brought to the form
\be
\label{2.50}
\hat{\vec{X}} = i |\vec{k}|^{1/2}\bs{E} \; \nabla_{\vec{k}} \, |\vec{k}|^{-1/2}{\bs{E}}^{-1} =
i |\vec{k}|^{1/2}\bs{E} \; \nabla_{\vec{k}} \, |\vec{k}|^{-1/2}{\bs{E}}^{\rm T}
\ee
where $3 \times 3$ unitary matrix $\bs{E}$ is defined as \eqref{2.34} by any orthonormal basis $\left(\vec{E}_1(\vec{k}), \vec{E}_2(\vec{k}), \vec{E}_3(\vec{k})   \right)$ on $\rtt=\mathbb{R}^3 \setminus \{ (0,0,k_3) \in \mathbb{R}^3 : k_3 \geq 0 \}$
satisfying \eqref{2.32}.} In that sense we say about uniqueness of the photon position operator with commuting components.
\begin{rmrk}
From the results given in \cite{hawton1, hawton5, dobrski} and those contained in the present section we quickly infer that one can start with other dense open submanifolds of the momentum space $\mathbb{R}^3$ admitting an orthonormal basis $\left(\vec{E}_1, \vec{E}_2, \vec{E}_3 \right)$ satisfying the conditions (\ref{2.32}). Then the operator of the form (\ref{2.48}) constructed by the use of that basis is determined by the unitary transformation of the operator constructed with the use of the triad $\left(\vec{E}_1(\vec{k}), \vec{E}_2(\vec{k}), \vec{E}_3(\vec{k})   \right)$ on $\rtt$ according to (\ref{2.48}).
\end{rmrk}


\section{Photon position operator with commuting components as a flat connection in a vector bundle}
\label{sec3}

In our previous work \cite{dobrski} we developed interpretation of the Hawton photon position operator and other photon position operators with commuting components in the momentum representation as defined by a flat connection $\hat{\vec{D}}, \; \hat{\vec{X}} =i \hat{\vec{D}}$ in some tangent bundle over a dense submanifold of the Euclidean space ${\mathbb R}^3$. 
The requirement of flatness 
\begin{equation}
\label{dflat}
[\hat{D}_i, \hat{D}_j]=0
\end{equation}
is a geometric counterpart of commutativity condition for components of photon position operator.
Connection $\hat{\vec{D}}$ obeys property that $2$--planes orthogonal to the momentum are transported parallel and  $\hat{\vec{D}}$ is an anti--Hermitian operator with respect to the Białynicki--Birula scalar product
\eqref{2.5}. As it was shown, flat connections determining the photon position operators are defined by absolutely parallel orthogonal bases of tangent vector fields $\left( \vec{e}_1(\vec{k}), \vec{e}_2(\vec{k}), \vec{e}_3(\vec{k})\right)$ such that 
$\vec{e}_{1}(\vec{k}), \vec{e}_{2}(\vec{k}) \, \bot \,\vec{k}.$ Therefore, these connections are particular examples of well known in differential geometry the {\it Weitzenb{\"o}ck connection}. Since a $2$--sphere is not a parallizable  manifold, in \cite{dobrski} we were forced to deal with some open dense submanifolds of ${\mathbb R}^3$ instead of the entire momentum space ${\mathbb R}^3.$
Thus we considered the following submanifolds:
${\mathbb R}^3 \setminus \{(0,0,k_3)\}_{k_3 \geq 0} =: \rtt$, ${\mathbb R}^3 \setminus \{(0,0,k_3)\}_{k_3 \in {\mathbb R}}$,
and ${\mathbb R}^3 \setminus \{(0,0,k_3)\}_{k_3 \leq 0}$. 


The original Hawton operator \cite{hawton} as well as several other photon position operators \cite{hawton1,dobrski} reveal string singularities. As it has been shown in \cite{hawton1} one can avoid these singularities by suitable gauge transformation of the form (\ref{2.26}), but then the gauge function $B \ofk$ is not globally univalent and, consequently, some discontinuites in the global expression for the photon position operator appear. This feature resembles very much the case of the vector potential of the Dirac magnetic monopole. Therefore it seems reasonable to use in our case the fibre bundle theory analogously as T.~T.~Wu and C.~N.~Yang \cite{wu,wu2} used it to understand the structure of the magnetic monopole field.  
\subsection{Trivial bundle over $M$}
Let us introduce the set
\begin{equation}
P=\left\{ \left( \vec{k},\bs{E} \ofk, \bs{\tPsi} \ofk \right)
   : \vec{k} \in \mathbb{R}^3 \setminus \{(0,0,0)\}  \right\}
\end{equation}
where $\bs{E} \ofk$ is defined by (\ref{2.34}) for $(\vec{E}_1 \ofk, \vec{E}_2 \ofk, \vec{E}_3 \ofk)$ satisfying conditions (\ref{2.32}) and with $\bs{\tPsi} \ofk$ fulfilling the transversality requirement (\ref{2.4}). One establishes an equivalence relation in $P$ according to
\begin{equation}
\left(\vec{k}, \bs{E} \ofk, \bs{\tPsi} \ofk \right) \sim \left(\vec{k}'', \bs{E''} (\vec{k}''), \bs{\tPsi}'' (\vec{k}'') \right) \iff \vec{k} = \vec{k}'' \quad \mathrm{and} \quad \bs{\tPsi}'' \ofk = \bs{E}'' \bs{E}^{-1} \bs{\tPsi} \ofk
\end{equation}
We are going to show that the quotient set $W=P/{\sim}$ has a natural structure of a trivial vector bundle endowed with a flat connection \cite{kobayashi, sulanke, maurin, husemoller}. Elements of $W$ are equivalence classes $\left[\left(\vec{k},\bs{E} \ofk, \bs{\tPsi} \ofk\right)\right]$ and the base manifold is the space $M= \mathbb{R}^3 \setminus \{(0,0,0)\}$ with the obvious projection $\pi : W \to M$ 
\begin{equation}
\pi\left(\left[\left(\vec{k},\bs{E} \ofk, \bs{\tPsi} \ofk\right)\right]\right)=\vec{k}
\end{equation}
The fibre over $\vec{k}$, $W_{\vec{k}} = \pi^{-1} \ofk$, is a two dimensional complex vector space with a well-defined linear operation
\begin{equation}
\alpha \left[\left(\vec{k},\bs{E} \ofk, \bs{\tPsi}_1 \ofk\right)\right] + \beta \left[\left(\vec{k},\bs{E} \ofk, \bs{\tPsi}_2 \ofk\right)\right] = 
\left[\left(\vec{k},\bs{E} \ofk, \alpha \bs{\tPsi}_1 \ofk + \beta \bs{\tPsi}_2 \ofk\right)\right]
\end{equation}
for $\alpha, \beta \in \mathbb{C}$. 
To introduce a differential structure and local trivializations one can make use of orthonormal bases considered in our previous work \cite{dobrski}. Let $(\vec{E}^S_1 \ofk$, $\vec{E}^S_2 \ofk)$, $\vec{E}^S_3 \ofk)$ be the \emph{south} basis defined on $\mathbb{R}^3_S =\rtt = \mathbb{R}^3 \setminus \{ (0,0,k_3) \in \mathbb{R}^3: k_3 \geq 0 \}$), and let $(\vec{E}^{N}_1 \ofk$, $\vec{E}^{N}_2 \ofk$, $\vec{E}^{N}_3 \ofk)$ be the \emph{north} basis defined on $\mathbb{R}^3_N:=\mathbb{R}^3 \setminus \{ {(0,0,k_3) \in \mathbb{R}^3}: {k_3 \leq 0} \}$. The matrices (\ref{2.34}) corresponding to these bases are $\bs{E}^S \ofk$ and $\bs{E}^N \ofk$ respectively. The functions
\begin{equation}
\bs{\tPhi}^S_A \ofk = \begin{pmatrix}
E^S_{A 1} \ofk\\ E^S_{A 2} \ofk \\ E^S_{A 3} \ofk
\end{pmatrix}
\quad 
\mathrm{and} 
\quad 
\bs{\tPhi}^N_A \ofk = \begin{pmatrix}
E^{N}_{A 1} \ofk\\ E^{N}_{A 2} \ofk \\ E^{N}_{A 3} \ofk
\end{pmatrix}
\quad
\mathrm{for} \quad A=1,2
\end{equation}
define local sections of $W$
\begin{equation}
t_A^S \ofk =\left[\left(\vec{k}, \bs{E}^S \ofk , \bs{\tPhi}^S_A \ofk \right)\right]
\quad \mathrm{and} \quad 
t_A^N \ofk =\left[\left(\vec{k}, \bs{E}^N \ofk , \bs{\tPhi}^N_A \ofk \right)\right]
\end{equation}
For a fixed vector $\vec{k}$ the elements $t^S_1 \ofk$, $t^S_2 \ofk$ (as well as $t^N_1 \ofk$, $t^N_2 \ofk$) form a basis of the fibre $W_{\vec{k}}$ over $\vec{k}$. We demand that the two bijective mappings 
\begin{equation}
\phi_S : \pi^{-1}(\mathbb{R}^3_S) \to \mathbb{R}^3_S \times \mathbb{C}^2  \quad \mathrm{and}  \quad \phi_N : \pi^{-1}(\mathbb{R}^3_N) \to \mathbb{R}^3_N \times \mathbb{C}^2 
\end{equation}
determined by the inverse formulas
\begin{equation}
\phi_{S}^{-1} (\vec{k},(f^1,f^2))=f^A t^S_A \ofk
\quad \mathrm{and} \quad
\phi_{N}^{-1} (\vec{k},(f^1,f^2))=f^A t^N_A \ofk
\end{equation}
(summation over $A=1,2$) are smooth local trivializations of $W$. Observe that for every $\vec{k} \in \mathbb{R}^3_S \cap \mathbb{R}^3_N$ 
\begin{equation}
\bs{E}^N {\bs{E}^S}^{-1} \bs{\tPhi}^S_A \ofk = \bs{E}^N {\bs{E}^S}^{T} \begin{pmatrix}
E^S_{A 1} \ofk\\ E^S_{A 2} \ofk \\ E^S_{A 3} \ofk
\end{pmatrix}
= 
\begin{pmatrix}
E^{N}_{A 1} \ofk\\ E^{N}_{A 2} \ofk \\ E^{N}_{A 3} \ofk
\end{pmatrix} = \bs{\tPhi}^{N}_A \ofk
\end{equation}
and consequently $t_A^S \ofk = t_A^N \ofk$ for $A=1,2$. This means that on the intersection $ \mathbb{R}^3_S \cap \mathbb{R}^3_N$ the mapping $\phi_N \circ \phi_S^{-1}$ acts as identity $(\vec{k},(f^1,f^2)) \mapsto (\vec{k},(f^1,f^2))$, thus the bundle $W$ is trivial. Moreover, this also means that sections $t_A^S \ofk$ and $t_A^N \ofk$ mutually prolong each other to the global sections $t_A \ofk$, $\vec{k} \in M$, $A=1,2$. In turn we can consider the global trivialization $\phi : W \to M \times \mathbb{C}^2$
\begin{equation}
\label{globtriv}
\phi^{-1} (\vec{k},(f^1,f^2))=f^A t_A \ofk
\end{equation}

The fiberwise scalar product in $W_{\vec{k}}$ given by
\begin{equation}
\label{fibsp}
\left( \left[\left(\vec{k},\bs{E} \ofk, \bs{\tPsi}_1 \ofk\right)\right] , \left[\left(\vec{k},\bs{E} \ofk, \bs{\tPsi}_2 \ofk\right)\right] \right)_{\vec{k}}
= \bs{\tPsi}^{\dagger}_1 \ofk \bs{\tPsi}_2 \ofk
\end{equation}
is properly defined since it does not depend on particular choice of $\bs{E} \ofk$. Notice that the scalar product $\left( t_A \ofk, t_B \ofk \right)_{\vec{k}} = \delta_{AB}$. For sections $s \ofk$, $u \ofk$ of $W,$ their scalar product (the counterpart of Białynicki--Birula scalar product (\ref{2.5})) can be taken as
\begin{equation}
\label{Wscprod}
\langle s  | u \rangle = \int_M \volel \left(s \ofk, u \ofk \right)_{\vec{k}}
\end{equation}

The connection $\nabla : \Gamma(W) \to \Gamma(W \otimes \Lambda^1)$ in $W$, where $\Gamma$ denotes the space of smooth sections of the corresponding bundle, can be introduced using $\hat{D}^N_j$ and $\hat{D}^S_j$ operators defined in terms of $\bs{E}^N$ and $\bs{E}^S$ matrices respectively. Indeed, if we put for $\vec{k} \in \mathbb{R}^3_N$
\begin{align}
\nonumber
\nabla \left[\left(\vec{k},\bs{E}^N \ofk, \bs{\tPsi} \ofk\right)\right] &=
\nabla_j \left[\left(\vec{k},\bs{E}^N \ofk, \bs{\tPsi} \ofk\right)\right] dk_j
=
\left[\left(\vec{k},\bs{E}^N \ofk, \hat{D}^N_j \bs{\tPsi} \ofk\right)\right]dk_j \\
&= \left[\left(\vec{k},\bs{E}^N \ofk, 
 |\vec{k}|^{1/2}  \bs{E}^N(\vec{k}) \frac{\partial}{\partial k_j}
 |\vec{k}|^{- 1/2} {\bs{E}^N}^{-1}(\vec{k}) \bs{\tPsi}(\vec{k})
\right)\right]dk_j
\end{align}
and analogously for $\vec{k} \in \mathbb{R}^3_S$ by means of $D^S_j$ and $\bs{E}^S$, then the definition is consistent on $\mathbb{R}^3_S \cap \mathbb{R}^3_N$ since for $\bs{\tPsi}'' \ofk = {\bs{E}^S} \ofk {\bs{E}^N}^{-1} \ofk  \bs{\tPsi} \ofk $ one obtains
\begin{align}
\hat{D}^S_j \bs{\tPsi}''(\vec{k})& = 
 |\vec{k}|^{1/2}  \bs{E}^S (\vec{k}) \frac{\partial}{\partial k_j}
 |\vec{k}|^{- 1/2} {\bs{E}^S}^{-1}(\vec{k}) \bs{\tPsi}''(\vec{k}) \nonumber \\
& = 
 |\vec{k}|^{1/2}  \bs{E}^S (\vec{k}){\bs{E}^N}^{-1}(\vec{k}) {\bs{E}^N}(\vec{k}) \frac{\partial}{\partial k_j}
 |\vec{k}|^{- 1/2} {\bs{E}^S}^{-1}(\vec{k}) {\bs{E}^S}(\vec{k}) {\bs{E}^N}^{-1}(\vec{k})  \bs{\tPsi}(\vec{k}) \nonumber \\
& = \bs{E}^S (\vec{k}) {\bs{E}^N}^{-1}(\vec{k}) \hat{D}^N_j \bs{\tPsi}(\vec{k}) 
\end{align}
The property (\ref{dflat}) of $\hat{D}^N_j$ and $\hat{D}^S_j$ ensures that the connection $\nabla$ is indeed flat. It is also anti-Hermitian with respect to the scalar product (\ref{Wscprod}) and hence the photon position operator $\hat{X}_j = i\nabla_j$ is still Hermitian. 

Finally, it is possible to define a helicity operator in the bundle $W$. Let $\bs{E}_{\bot}$ be a $3 \times 2$ matrix created from first two columns of $\bs{E}$ (thus, it contains components of vectors $\vec{E}_1$ and $\vec{E}_2$). For $\bs{\Sigma}$ given by (\ref{2.2}) and an arbitrary specific choice of $\bs{E}$ it can be verified by direct calculation that
\begin{equation}
\label{SigmForm} 
\bs{\Sigma} = \bs{E}_{\bot} \sigma_2 \bs{E}^{T}_{\bot}
\end{equation}
with 
\begin{equation}
\sigma_2 = \begin{pmatrix}0 & -i\\ i & 0\end{pmatrix}
\end{equation}
If $\bs{E}^{''}_{\bot}$ is related to $\bs{E}_{\bot}$ by $\bs{E}^{''}_{\bot}=\bs{E}_{\bot} \bs{U}_{\bot}$ for
\begin{equation}
\bs{U}_\bot = \begin{pmatrix}a & b\\ -b & a\end{pmatrix}
\end{equation}
with $a^2 +b^2 =1$, then, since $\bs{U}_\bot \sigma_2 \bs{U}^{T}_\bot = \sigma_2$, it follows that $\bs{E}_{\bot} \sigma_2 \bs{E}^{T}_{\bot} = \bs{E}^{''}_{\bot} \sigma_2 {\bs{E}^{''}_{\bot}}^T$. Thus the formula (\ref{SigmForm}) is universal. Using this observation and performing straightforward calculation one obtains
\begin{equation}
\bs{\Sigma} \bs{E}^{''} \bs{E}^{-1} =  \bs{E}^{''} \bs{E}^{-1} \bs{\Sigma}
\end{equation}
In turn, it may be concluded that the helicity operator in $W$ given by
\begin{equation}
\bs{\Sigma} \left[\left(\vec{k},\bs{E} \ofk, \bs{\tPsi} \ofk\right)\right]
 = \left[\left(\vec{k},\bs{E} \ofk, \bs{\Sigma} \bs{\tPsi} \ofk\right)\right]
\end{equation}
is well defined.

\subsection{Quantum mechanics of photon in $M \times \mathbb{C}^2$}
Using the global trivialization $\phi$ of bundle $W$ defined by (\ref{globtriv})
one can transport quantum-mechanical structures to $M \times \mathbb{C}^2$. Let $\psi \ofk = \begin{pmatrix} \psi^1 \ofk \\ \psi^2 \ofk \end{pmatrix}$ and $\varphi \ofk = \begin{pmatrix} \varphi^1 \ofk \\ \varphi^2 \ofk \end{pmatrix}$ be sections of the trivial bundle $M \times \mathbb{C}^2$ and let $\phi^{-1}(\psi) \ofk = \phi^{-1}(\vec{k},(\psi^1 \ofk,\psi^2 \ofk))=\psi^A \ofk t_A \ofk$ and $\phi^{-1} (\varphi) \ofk = \phi^{-1}(\vec{k},(\varphi^1 \ofk,\varphi^2 \ofk))=\varphi^B \ofk t_B \ofk $ be the corresponding sections of $W$. The scalar product of sections $\psi$ and $\varphi$ can be obtained in terms of formula (\ref{Wscprod})
\begin{align}
\nonumber
\langle \psi | \varphi \rangle & = \langle \phi^{-1}(\psi) | \phi^{-1} (\varphi) \rangle  = \langle \psi^A t_A | \varphi^B t_B \rangle 
\\ & = \int_M \volel \left(\psi^A \ofk \right)^* \varphi^B \ofk \left(t_A \ofk, t_B \ofk \right)_{\vec{k}} =
\int_M \volel \psi^{\hrm} \ofk \varphi \ofk 
\end{align}

Now, let us consider a photon position operator acting on sections of $M \times \mathbb{C}^2$. First, one can observe that for $j=1,2,3$ and $A=1,2$
\begin{equation}
\label{nabont}
\nabla_j t_A = -\frac{k_j}{2 |\vec{k}|^2} t_A
\end{equation}
For the global trivialization $\phi$ given by (\ref{globtriv}), let $\phi_{\vec{k}} : W_{\vec{k}} \to \mathbb{C}^2$ be defined by $\phi(s) =(\pi(s),\phi_{\vec{k}} (s))$, where $\vec{k}=\pi(s)$ and $s \in W$. Then, for a section $\psi \in \Gamma(M \times \mathbb{C}^2)$ the photon position operator becomes 
\begin{align}
\nonumber
\hat{X}_j \psi \ofk & = \phi_{\vec{k}} \left(\hat{X}_j \phi^{-1}(\psi) \ofk\right) = \phi_{\vec{k}} \left(i \nabla_j \left(\psi^A \ofk t_A \ofk \right) \right) \\
\label{trivposop}
& = 
\phi_{\vec{k}} \left( i\left( \frac{\partial \psi^A}{\partial k_j} \ofk -\psi^A \ofk \frac{k_j}{2 |\vec{k}|^2}\right)t_A \ofk\right) =  i\left( \frac{\partial }{\partial k_j}-\frac{k_j}{2 |\vec{k}|^2}\right) \psi \ofk
\end{align}
Notice that this is the well-known position operator for spinless particle introduced by Newton and Wigner \cite{newton}.

The helicity operator in $M \times \mathbb{C}^2$ can be obtained in the similar manner. Using formula (\ref{SigmForm}) one can easily calculate that
$\bs{\Sigma}\bs{\tPhi}^S_A \ofk = \bs{\tPhi}^S_B \ofk \tensor{{\sigma_2}}{^B_A}$. As the analogous result holds for the north basis it follows that
\begin{equation}
\bs{\Sigma} t_A \ofk = t_B \ofk \tensor{{\sigma_2}}{^B_A}
\end{equation}
Thus, when $\bs{\Sigma}$ acts on a section $\psi \in \Gamma(M \times \mathbb{C}^2)$ it produces
\begin{align}
\nonumber
\bs{\Sigma} \psi \ofk & = 
\phi_{\vec{k}} \left( \bs{\Sigma} \phi^{-1}(\psi) \ofk \right)=
\phi_{\vec{k}} \left( \psi^A \ofk \bs{\Sigma} t_A \ofk\right) 
= \phi_{\vec{k}} \left(  t_B \ofk \tensor{{\sigma_2}}{^B_A} \psi^A \ofk \right) 
\\
& = \sigma_2 \psi \ofk
\end{align}
This means that the helicity operator becomes the second Pauli matrix $\sigma_2$ in the considered case.


Finally, using the eigenvectors of photon position and helicity operators obtained for $\rtt$, one can look for their $M \times \mathbb{C}^2$ counterparts. These eigenvectors read (compare \cite{dobrski})
\begin{equation}
\bs{\tPsi}_{\vec{X}, \pm 1}\ofk  = \frac{1}{\sqrt{2}} \left( \bs{E}_1 \ofk \pm i \bs{E}_2 \ofk \right) \kn^{\frac{1}{2}} e^{- i \vec{k} \cdot \vec{X}}  
\end{equation}
where $\bs{E}_1 \ofk$ and $\bs{E}_2 \ofk$ are columns of components of vectors $\vec{E}_1 
\ofk$ and $\vec{E}_2 \ofk$ respectively. They satisfy
\begin{align}
\hat{\vec{X}} \bs{\tPsi}_{\vec{X}, \pm 1} &= \vec{X} \bs{\tPsi}_{\vec{X}, \pm 1} & \bs \Sigma \bs{\tPsi}_{\vec{X}, \pm 1} &=  \pm  \bs{\tPsi}_{\vec{X}, \pm 1}
\end{align} 
When transported to $M \times \mathbb{C}^2$ the eigenvectors take the following simple form
\begin{equation}
\psi_{\vec{X}, \pm 1}\ofk= \frac{1}{\sqrt{2}}   \begin{pmatrix}1 \\ \pm i \end{pmatrix} \kn^{\frac{1}{2}}  e^{- i \vec{k} \cdot \vec{X}}   
\end{equation}
%
%
Of course, one can use some other basis of global sections to establish trivialization isomorphism. The most natural seems to be the basis defined by complex polarization vectors
\begin{align}
t'_1 \ofk & = \frac{1}{\sqrt{2}} \left(t_1 \ofk + i t_2 \ofk \right) & t'_2 \ofk& = \frac{1}{\sqrt{2}} \left( t_1 \ofk - i t_2 \ofk \right)
\end{align} 
Then, one quickly finds that the operator $\hat{\vec{X}}$ in this new basis acts, \emph{mutatis mutandi}, according to (\ref{nabont}) and (\ref{trivposop}), while the helicity operator $\bs{\Sigma}$ is represented by the third Pauli matrix 
\begin{equation}
\sigma_3 = \begin{pmatrix}1 & 0\\ 0 & -1\end{pmatrix}
\end{equation}
The eigenfunctions of $\hat{\vec{X}}$  and $\bs{\Sigma}$ read now
\begin{align}
\psi^+_{\vec{X}}\ofk &= \begin{pmatrix}1 \\ 0 \end{pmatrix} \kn^{\frac{1}{2}}  e^{- i \vec{k} \cdot \vec{X}} & 
\psi^-_{\vec{X}}\ofk &= \begin{pmatrix}0 \\ 1 \end{pmatrix} \kn^{\frac{1}{2}}  e^{- i \vec{k} \cdot \vec{X}} 
\end{align}

%
%
We conclude that using our vector bundle construction one can straightforwardly express the photon quantum mechanics (corresponding to the Hawton position operator) using  operators acting on two-component wave functions defined on $\mathbb{R}^3 \setminus \{(0,0,0)\}$.
The remarkable advantage of this approach is the possibility of defining the photon position operator with commuting components which does not reveal any singularities. 


\section{Concluding remarks}
\label{sec4}
 
In this work we  present a general form of the photon position operator with commuting components. For its construction we postulate that, besides the mutual commutativity of its components,
the canonical operator of the photon momentum is given and that the corresponding position operator fulfils the following assumptions: it commutes with the photon helicity operators, its action on a photon wave function produce another photon wave function and that it is Hermitian with respect to the Bialynicki – Birula scalar product.

It is shown that up to a unitary transformation satisfying the transversality condition, other proposals of photon position operators with commuting components, as the one given by M. Hawton, are special cases of the general photon position operator given in Eq. \eqref{2.50}. 
Nevertheless, all proposed photon position operators with commuting components exhibit string singularities or some discontinuities which cannot be avoided by a global gauge transformation of the form (\ref{2.26}) with univalent function $B \ofk$. This situation is quite similar to that of the Dirac magnetic monopole. Therefore we apply the fibre bundle method of T.~T.~Wu and C.~N.~Yang \cite{wu,wu2} and assume that the photon wave functions
are sections of appropriately constructed vector bundle over $M=\mathbb{R}^{3} \setminus \{(0,0,0)\}$. This then allows to define the photon position operator in terms of a connection in this bundle. Finally we are able to transport the whole construction to $M \times \mathbb{C}^2$ and obtain a version of photon quantum mechanics with wave functions being dependent solely on physical, commuting degrees of freedom -- nonzero momentum and $\pm 1$ helicity.

\section*{Acknowledgments}
The work of F.J.T.\ was partially supported by SNI-México, COFAA-IPN and by Secretaría de Investigación y Posgrado del IPN Grant No.\ 20221484.

\end{document}